\def\mathfont#1{\ifmmode{#1}\else{$#1$}\fi} 
\def\lae{\mathrel{<\kern-1.0em\lower0.9ex\hbox{$\sim$}}}  
\def\gae{\mathrel{>\kern-1.0em\lower0.9ex\hbox{$\sim$}}}  

\def\msun{\ifmmode{\ {\rm M}_\odot}\else{$ {\rm M}_\odot$}\fi}  
\def\msunyr{\ifmmode{\msun \ {\rm yr}^{-1}}\else{$\msun \ {\rm 
yr}^{-1}$}\fi}  

\input psfig.sty

\documentclass[12pt,preprint]{aastex}

\slugcomment{Accepted for publication in the ApJ Letters.}

\shorttitle{Cavities in Abell 2597}
\shortauthors{McNamara et al.}

\begin{document}


\title{Discovery of Ghost Cavities in Abell 2597's X-ray Atmosphere}



\author{B. R. McNamara\altaffilmark{1,6},
M. W. Wise\altaffilmark{2},
P. E. J. Nulsen\altaffilmark{3},
L. P. David\altaffilmark{6},
C. L. Carilli\altaffilmark{7},
C. L. Sarazin\altaffilmark{5},
C. P. O'Dea\altaffilmark{4},
J. Houck\altaffilmark{2},
M. Donahue\altaffilmark{4},
S. Baum\altaffilmark{4},
M. Voit\altaffilmark{4},
R. W. O'Connell\altaffilmark{5},
A. Koekemoer\altaffilmark{4}}

\altaffiltext{1}{Ohio University, Dept. of Physics \& Astronomy, Athens,
OH 45701}
\altaffiltext{2}{Massachusetts Institute of Technology, Center for Space Research, 70 Vassar St., Cambridge, MA 02139}
\altaffiltext{3}{Engineering Physics, University of Wollongong, Wollongong NSW 2522, Australia }
\altaffiltext{4}{Space Telescope Science Institute, 3700 San Martin Drive, Baltimore, MD 21218}
\altaffiltext{5}{Dept. of Astronomy, University of Virginia, P.O. Box 3818,
Charlottesville, VA 22903}
\altaffiltext{6}{Harvard-Smithsonian Center for Astrophysics, 60 Garden St., Cambridge, MA 02138}
\altaffiltext{7}{National Radio Astronomy Observatory, Socorro, NM}

\begin{abstract}
A Chandra image of the central 100 kpc of
the Abell 2597 cluster of galaxies shows bright, irregular, X-ray emission
within the central dominant cluster galaxy (CDG),
and two low surface brightness cavities located 30 kpc from
the CDG's nucleus.  Unlike the cavities commonly seen in other clusters,
Abell 2597's ``ghost'' cavities are not coincident with the bright central
radio source.  Instead, they appear to be associated with faint,
extended  radio emission seen in a deep VLA radio map.
We interpret the ghost cavities as buoyantly-rising relics of
a radio outburst that occurred between 50--100 Myr ago.
The demography of cavities in the few clusters studied thus far 
shows that galactic radio sources experience 
recurrent outbursts on a $\sim 100$ Myr timescale.  
Over the lifetime of a cluster, ghost cavities emerging
from CDGs deposit $\gae 10^{59-61}$ erg of energy into the 
intracluster medium.  If a significant fraction of this energy is 
deposited as magnetic field,
it would account for the high field strengths in the cooling flow
regions of clusters.  The similarity between the central cooling 
time of the keV gas and the radio cycling timescale
suggests that feedback between cooling gas and the radio
source may be retarding or quenching the cooling flow.

\end{abstract}




\section{Introduction}

Early Chandra images of galaxy clusters have shown that 
the X-ray emitting gas in their centers is bright and irregularly 
structured, and that much of this structure is 
associated with powerful radio sources.  The radio sources in the
Hydra A (McNamara et al. 2000),  Perseus (Fabian et al.
2000), and Abell 2052 (Blanton et al. 2001) clusters 
appear to have pushed aside the keV 
gas leaving low surface brightness cavities in the gas.  
The cavities in Hydra A, Perseus, and Abell 2052 are filled
with bright radio emission and are confined by the pressure of
the surrounding keV gas. The cavities may be supported against collapse 
by pressure from relativistic particles,  magnetic fields, and/or
hot, thin thermal gas.  Since the cavities have a lower gas density
than their surroundings, they should behave
like bubbles in water and rise buoyantly in the intracluster 
medium (ICM; McNamara et al. 2000; Churazov et al. 2001).  

Using simulations of supersonic jets expanding 
into the ICM,  Clarke, Harris, \& Carilli (1997)
and Heinz, Reynolds, \& Begelman (1998) 
argued that the cavities seen in  {\it ROSAT} images of 
the ICM surrounding the Perseus and Cygnus A radio
sources (B\"ohringer et al. 1992; Carilli, Perley, \& Harris 1994)
were caused by strong shocks.   
In this instance, the X-ray emission from the rims surrounding the cavities
should be spectrally hard, and gas in the rim should have
higher entropy than the surrounding gas.  The initial Chandra
results for the Hydra A (McNamara et al. 2000; Nulsen et al. 2001),
Perseus (Fabian et al. 2000) and Abell 2052 (Blanton et al. 2001) clusters
were surprising, as the emission
from the rims of the cavities was among the softest in the clusters.
This implies that the radio lobes expanded gently into the
intracluster medium at roughly the sound speed 
in the keV gas (Reynolds, Heinz, \& Begelman 2001; David et al. 2000;
Nulsen et al. 2001). The rapidly growing number of cavities
found in giant elliptical galaxies (Finoguenov \& Jones 2000),
groups (e.g. Vrtilek et al. 2000) and 
cluster CDGs (e.g. Schindler et al. 2001) indicates that they are 
persistent features of these systems.

An intriguing and potentially significant Chandra discovery 
is the existence of cavities
in the keV gas that do not have a bright radio counterpart.
If such ``ghost'' cavities seen in Perseus (Fabian et al. 2000),
and in Abell 2597, discussed here and in McNamara et al. (2000), 
are generated  by radio sources, their properties would have significant consequences for our understanding of the life cycles of radio galaxies
and the origin and dispersal of magnetic fields in clusters and galaxies.
Here we discuss the remarkable properties
of the X-ray core of Abell 2597 and its ghost cavities.
Throughout this paper, we assume ${\rm H_0}=70~{\rm km ~s^{-1}~Mpc^{-1}}$,
$\Omega_{\rm M}=0.3$, $\Omega_\Lambda =0.7$, a luminosity distance of 374
Mpc, and 1 arcsec = 1.67 kpc.

\section{Observations \& Data Reduction}

Abell 2597 is an 
$L_{\rm x}=6.45\times 10^{44}~{\rm erg~sec}^{-1}$ ($2-10$ keV; David 
et al. 1993), richness class 0 cluster that
lies at redshift $z=0.083$.  The cluster possesses
a bright cusp of X-ray emission associated
with a cooling flow and the powerful radio source 
PKS 2322-122, both centered on the CDG (Sarazin et al. 1995).  

A 40 ksec Chandra exposure was taken of Abell 2597 on 2000 July 28.
The nucleus of the BCG was centered on node 0 of the ACIS S3
back-illuminated device.  The pointing was chosen to maximize the
spatial resolution and soft energy response available with Chandra
while avoiding placing the interesting central region on a node
boundary.  The observations were made in Faint, full-frame,
timed exposure mode, with the focal plane at temperature $-120$ C.
Grades 1, 5, and 7 were rejected in our analysis, as were data below 0.3 keV
and above 8.0 keV.  Unfortunately, the observations were compromised
by strong flares.  As a consequence, only 18.54 ksec of useful 
data were gathered.

\section{Comparison between the Central X-ray \& Radio Structures}

An adaptively smoothed X-ray image of the central 
$150\time 150$ arcsec centered on the CDG is shown in Figure 1.
The passband of this image is from $0.3-8.0$ keV.  The image
immediately reveals the gross structure in the 
inner arcmin or so of the cluster first seen in an earlier 
$ROSAT$ High Resolution  Imager  (HRI)
observation (Sarazin et al. 1995).  However, the fine structure
and particularly the 
surface brightness depressions located 18 arcsec to the south-west
and 16 arcsec to the north-east of the center were not
seen in the ROSAT image.  

The central structure was isolated 
by modeling and subtracting the smooth background cluster 
emission leaving the excess emission shown in Figure 2.  
This difference image was then adaptively smoothed, which revealed
structure shown above the $4\sigma$ significance level.  
In the bright regions, surface brightness variations of $50-60\%$
are present. The large surface brightness depressions to the north-east and south-west are two to three times fainter than the
surrounding regions, the depression to the south-west being deeper. 
In order to examine the relationship between the central radio
source and the structure in the gas, we have superposed
radio contours of Abell 2597 in Figure 2.  The
radio image was obtained with the VLA A configuration,
tuned to frequency 8.44 GHz (Sarazin et al. 1995).
The radio source is relatively small and has a steep spectrum
with a spectral index of $\simeq -1.5$ (O'Dea et al. 1994; Sarazin et al.
1995).  Its full extent from north
to south is only $5$ arcsec, or
roughly 8 kpc.   Although a great deal of structure
is seen surrounding the radio source, there is no 
evidence for cavities there in spite
of its powerful nature.  However, the central radio source is small
compared to the bright X-ray structure in the inner 40 kpc or so.  
Therefore, any cavities that may exist would be difficult to detect
due to emission from intervening  material along the line of sight.  

Unlike the cavities in Hydra A, Abell 2597's cavities are located at 
larger radii and are more than twice the size
of the central radio source.  After subtracting the background cluster,
the cavities appear to be more extensive than the impression given
in Figure 1. The cavity to the north-east
is roughly circular with a $\simeq 9$ arcsec diameter, 
corresponding to a linear diameter of 15 kpc.  The cavity to
the south-west is elliptically shaped with major and minor axes
$\simeq 14$ arcsec and $\simeq 9$ arcsec, or 23 kpc and 15 kpc
linear diameters, respectively.  The cavities are surrounded 
by shells of X-ray gas on most sides, but perhaps 
not at the outermost radii.  We divided the data into a soft
and hard band images with $0.5-1.5$ keV and $1.5-3.5$ keV passbands,
respectively, and we arrived at the following conclusions.
To within the accuracy of the data, 
1) the emission immediately adjacent to each cavity is generally no
harder or softer than its surroundings; 2) deep surface brightness 
depressions are present in both bands.  Therefore, there is no 
compelling evidence for heating by the agent that inflated the cavities, 
nor are the depressions likely to be caused by absorption.  

In order to determine whether faint radio emission
is associated with the cavities, 
radio observations of Abell 2597 were made with the
Very Large Array at 1.4 GHz on 2001 June 21. 
The total observing time was 12 hours, and the
array configuration was mixed between the 3 km and 10 km
configurations, leading to a synthesized beam FWHM = 
11 arcsec by  6 arcsec  with the major axis position angle = 90$^o$.
Standard amplitude and phase calibration were applied,
as well as self-calibration using sources in the field.
The dominant source in the field is the nucleus of
Abell 2597 itself (Figure 2), which has a peak surface brightness in
our 1.4 GHz image of 1.49 $\pm$ 0.03 Jy/beam. 
This bright source limits the sensitivity of the final image
to about 0.1 mJy/beam rms, implying a dynamic range of 15000.
The radio emission from Abell 2597 is marginally resolved with
these observations, with a total flux density of 1.82 Jy. 

In the context of studying the X-ray cavities, the most
interesting results from the radio observations 
are the extensions of the radio source in the vicinity of the
cavities, as can be seen in Figure 3.
These extensions are robust in the imaging process, and
are highly significant with respect to the noise on the image.
Unfortunately, the resolution of the image is inadequate to
make any firm conclusions concerning the morphology of
the extended emission in the vicinity of the cavities, 
only that such emission exists. 
This detection of radio emission and a future confirmation
with higher resolution are keys to understanding the
nature of ghost cavities (\S 5).

\section{Physical State of keV gas}

The radial distribution of surface brightness, temperature, 
electron density, and
pressure in the central region of Abell 2597 is shown in 
Figure 4.  The profiles were extracted from annular apertures centered
on the weak nuclear point source coincident with the radio core
(${\rm RA}=23~25~19.7$ , ${\rm Dec}=-12~07~27$, J2000).  The aperture
sizes were chosen to include roughly 1000 counts and 
5000 counts for surface brightness and temperature profiles shown in
Figures 4a and 4c respectively.  The density profile, Figure 4b, 
was constructed by deprojecting the surface brightness profile
assuming an emission measure appropriate for a 3 keV gas.
The temperature in each aperture was determined
by fitting an absorbed MEKAL single temperature 
model in  XSPEC with abundances fixed at 0.4 Solar, 
and a Galactic foreground column $N_{\rm H}=2.48\times 10^{20}~{\rm cm}^{-2}$. Only data from the
ACIS-S3 device were considered.

The temperature drops from 3.4 keV at 100 kpc to 
1.3 keV in the inner several kpc.  
Over this same region, the gas density and pressure rise dramatically,
reaching values of $0.07-0.08~{\rm cm}^{-3}$ and $2.5\times 10^{-10}~{\rm erg~cm^{-3}}$ in the inner few kpc, respectively.  The radiative cooling time of the keV gas in the
central 10 kpc region surrounding the radio source is
only $\approx 3\times 10^8$ years. Similar properties
are found in other cooling flow clusters, such as Hydra A
(McNamara et al. 2000; David et al. 2000), Abell 2052 (Blanton et al. 2001),
and Perseus (Fabian et al. 2000). 

\section{Origin \& Energy Content of the Ghost Cavities}

Without internal pressure support, 
the cavities would collapse on the sound crossing timescale of
$\sim 10^7$ yr.  Yet, the existence of ghost cavities in Abell 2597 and
in NGC 1275 (Fabian et al. 2000) beyond their radio lobes 
shows that they almost certainly persist 
much longer than $10^7$ yr, and therefore must have pressure support.
The gas density within the ghost cavities is much less than
the ambient density.  Therefore, they must be buoyant and have
risen outward from the nucleus to their current projected radii 
of $\simeq 30$ kpc.
The time required for the cavities to rise to this radius 
is roughly $10^{7.7-8}$ years (McNamara et al. 2000; Churazov et al. 2001).
This is much larger than the minimum age of the
central radio source $\sim 5\times 10^6$ yr (Sarazin et al. 1995),
so the ghost cavities are unlikely to be related directly to the
current nuclear radio episode.

Based on the properties of radio-bright cavities, 
it is reasonable to suppose that the ghost cavities were produced 
in a radio episode that predated the current one shown in Figure 2.
Their radio emission has faded presumably because they are no longer
being supplied with relativistic particles from the nucleus.
This hypothesis is supported by our detection of extended radio emission toward
the cavities and their locations along nearly the same
position angle as the central jets (Sarazin et al. 1995). 
A rough estimate of the minimum energy pressure in the
extended 1.4 GHz emission is considerably less than the ambient
pressure, as is found in other clusters.  This would suggest a departure
from the minimum energy condition, or a dominant pressure
contribution from low energy electrons or protons.
  
Assuming the cavities formed through the action of a radio
source, the  lower limit to the energy expended during their
formation is given by the  $PdV$ work done on the
surrounding gas.  The gas pressure at the radius of the
cavities is $\simeq 2.1\times 10^{-10}~{\rm erg~cm^{-3}}$.
Assuming the cavities are projected spheroids with volumes
of $\approx 5\times 10^{67}~{\rm cm}^{3}$ and  
$\approx 1\times 10^{68}~{\rm cm}^{3}$
for the north-east and south-west cavities, respectively,
the minimum energy corresponds to $\sim 3.4\times 10^{58}~{\rm erg}$.  
The figure could be $2-3$ times larger if one includes
the energy of thermal and relativistic gas
in the cavities.   This energy is more than 
an order of magnitude larger than the total mechanical energy 
of the central radio source (Sarazin et al. 1995).
Therefore, if the ghost cavities originated in an earlier radio
event, it would have been as powerful in the past as is Hydra A.

\section{Discussion}

The existence of ghosts cavities and their likely association 
with radio sources implies directly that  powerful radio sources 
occur in repeated outbursts.  This is consistent 
with the fact that more than 70\% of CDGs in clusters with bright 
X-ray cusps--cooling flows--harbor powerful
radio sources, while less than 20\% of CDGs in non cooling flow 
clusters are radio-bright (Burns 1990).  It would seem then that
radio sources in cooling flow CDGs have the shortest duty cycles
among the elliptical galaxies. 
The time required for cavities to rise to their observed
locations implies radio cycling every $50-100$ Myr.
The alternative interpretation of the high incidence of radio emission, 
that radio sources persist on Gyr timescales,
appears to be inconsistent with the existence of ghost cavities.

The cooling time of the central keV gas, $\approx 3\times 10^8$ yr,
is interestingly close to the radio cycling timescale.  This implies
that feedback between radio heating and radiative cooling may 
be operating there (Tucker \& David 1997; Soker et al. 2001).  
The energy deposited by cavities into the ICM in the form of 
magnetic fields, cosmic rays, and heat over the life of the 
cluster is $\gae 10^{60-61}$ erg, assuming the CDG produces 
between $10-100$ bubbles over its lifetime. 
This is roughly equivalent to the cooling luminosity 
of a $\approx 50 \msunyr$ cooling flow in the center of the cluster.  
While this energy could reduce and possibly quench a
small cooling flow, there is scant evidence for direct
heating by radio sources 
(McNamara et al. 2000; Fabian et al. 2000; David et al. 2001).
However, bulk lifting of cooling material out of 
cluster cores where it will expand, mix with
ambient gas, and cool less efficiently can assist in reducing the deposition
of cooled gas without the direct introduction of heat.
(David et al. 2001; Nulsen et al. 2001).

Finally, clusters are
magnetized  (Clarke, Kronberg, \& B\"ohringer 2001; Kronberg et al. 2001), 
and cavities emerging from the CDGs and normal elliptical galaxies
in clusters may be vessels 
that transport magnetic fields from galaxy nuclei to the intracluster medium.  
If a significant fraction of the
$10^{60-61}$ erg of energy emerging from BCGs alone were deposited as 
magnetic field in the inner 100 kpc of clusters, the implied field strengths 
of $\sim 5-50 \mu$G  would be consistent with
the field strengths observed in the cores of cooling flow
clusters (Ge \& Owen 1993).

\acknowledgments

B. R. M. thanks Liz Blanton for stimulating discussions.
This research was supported by LTSA grant NAG5-11025
and Chandra General Observer award GO0-1078A.
The National Radio Astronomy Observatory is
operated by Associated Universities, Inc., under a cooperative
agreement with the National Science Foundation.




\begin{figure}[h]
\hbox{
\hspace{.0in}
\psfig{figure=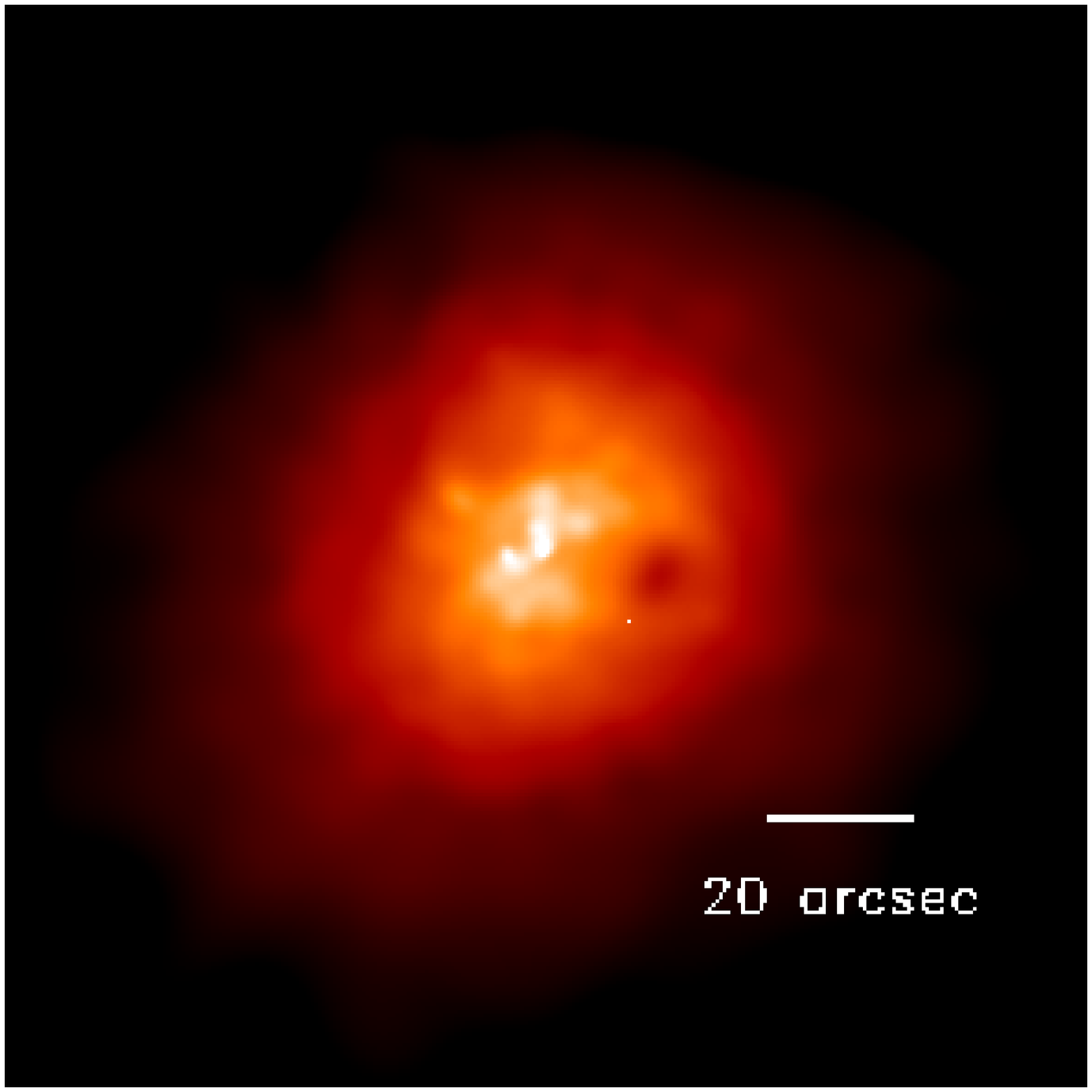,angle=90,height=5.0in,width=5.0in}
}
\begin{minipage}[h]{6.0truein}
Figure 1: Broadband smoothed X-ray image of Abell 2597. The surface brightness
is irregular in the central 40 arcsec.  The surface brightness
depressions associated with the cavities are seen 18 arcsec
to the south-west and north-east of center.  North is at top; east is at left.
\end{minipage}
\end{figure}

\begin{figure}[h]
\hbox{
\hspace{.0in}
\psfig{figure=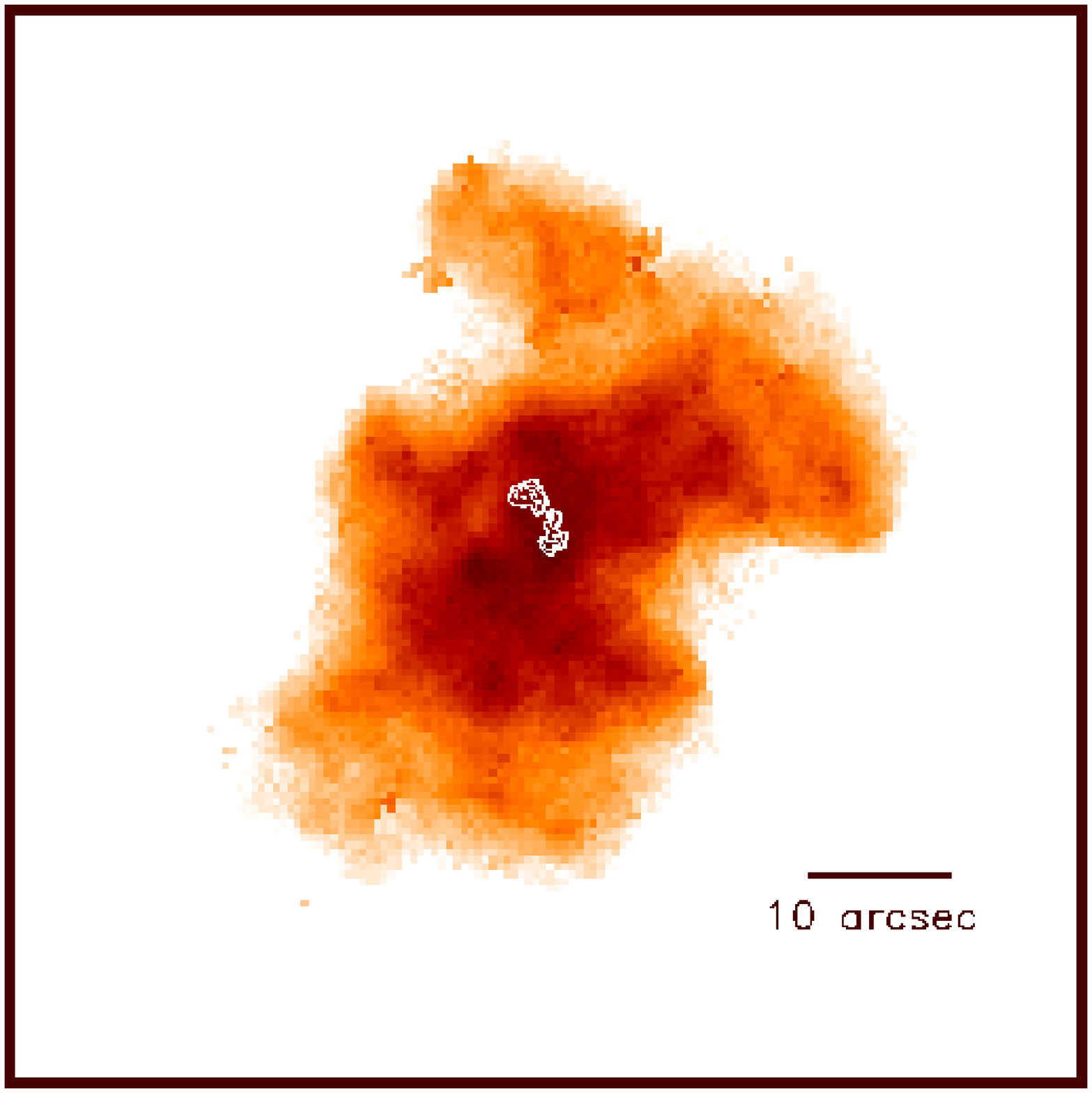,angle=90,height=5.0in,width=5.0in}
}
\begin{minipage}[h]{6.0truein}
Figure 2: Expanded view  of the central region of Abell 2597 after
subtracting a smooth background cluster model. The 8.44 GHz radio contours are
superposed. The cavities are
seen as indentations in the bright emission. North is at top; east is left.
\end{minipage}
\end{figure}

\clearpage
\begin{figure}[h]
\hbox{
\hspace{.0in}
\psfig{figure=brmfig3.ps,height=5.0in,width=5.0in,angle=-90}
}
\begin{minipage}[h]{6.0truein}
Figure 3: The VLA 1.4 GHz image of Abell 2597 at 
11 by 6 arcsec  resolution (oval contours), 
with the major axis position angle = 90$^o$. 
The first five contour levels are a geometric progression
in the square root of two, with the first level being
0.9 mJy/beam. The higher contours are a geometric progression
by factors of two.  The boxes correspond roughly to the positions
and sizes of the X-ray cavities.  The faint contours extending
south of center are data artifacts.
\end{minipage}
\end{figure}

\begin{figure}[h]
\hbox{
\hspace{.0in}
\psfig{figure=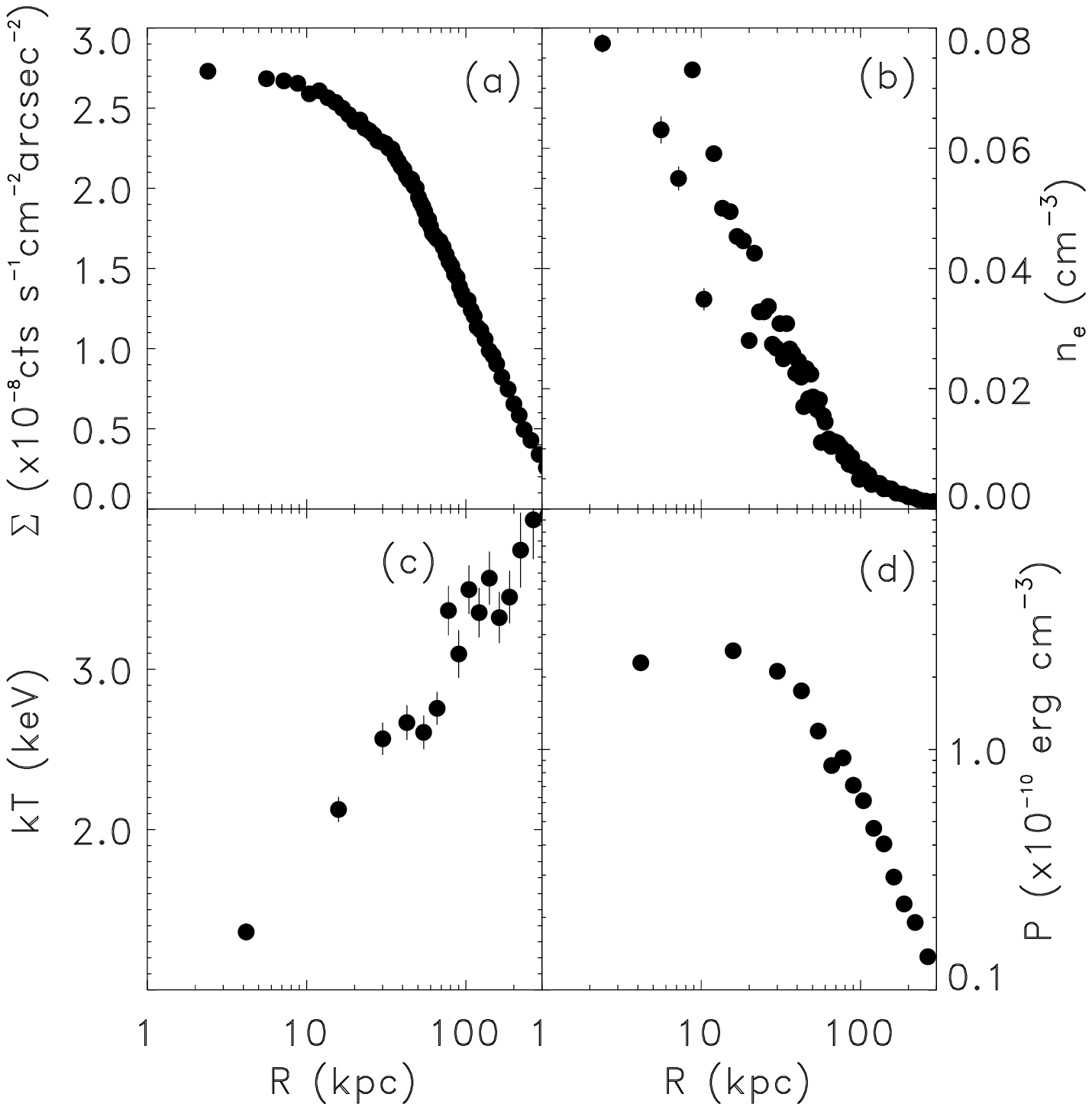}
}
\begin{minipage}[h]{6.0truein}
Figure 4: Radial variation of (a) surface brightness, 
(b) electron density,
(c) temperature, (d) pressure.
\end{minipage}
\end{figure}

\end{document}